\documentclass[12pt]{iopart}
\usepackage{amsfonts}
\usepackage{iopams}
\usepackage{graphicx}

\begin{document}

\title{Experimental realization of a quantum game on a one-way quantum
computer}
\author{Robert Prevedel$^1$, Andr\'{e} Stefanov$^{1,2}$, Philip Walther$^{3}$
and Anton Zeilinger$^{1,2}$}

\begin{abstract}
We report the first demonstration of a quantum game on an
all-optical one-way quantum computer. Following a recent
theoretical proposal we implement a quantum version of Prisoner's
Dilemma, where the quantum circuit is realized by a 4-qubit
box-cluster configuration and the player's local strategies by
measurements performed on the physical qubits of the cluster.
This demonstration underlines the strength and versatility of the
one-way model and we expect that this will trigger further
interest in designing quantum protocols and algorithms to be
tested in state-of-the-art cluster resources.
\end{abstract}

\address{$^1$Faculty of Physics, University of Vienna, Boltzmanngasse 5, A-1090 Vienna, Austria\\
$^2$Institute for Quantum Optics and Quantum Information (IQOQI),
Austrian Academy of Sciences, Boltzmanngasse 3, A-1090 Vienna,
Austria\\
$^{3}$Physics Department, Harvard University, Cambridge,
Massachusetts 02138, USA\\} \ead{zeilinger-office@quantum.at}

 \maketitle


\section{Introduction}

In the past, classical game theory has been extensively used to study
problems such as stock market development, human as well as animal behavior
or even the evolution of viruses at the microbiological level \cite%
{Myerson1991,Fudenberg1991,Turner1999}. Quantum versions \cite%
{Meyer1999,Eisert1999,Benjamin2001} of existing games offer additional
strategies to the players - and resolve dilemmas that occur in the classical
versions. As it is possible to recast any algorithm (classical or quantum)
as a game characterized by strategies and rules, it is reasonable to believe
that the quantum mechanical formulation of existing games can also be
helpful in gathering a deeper understanding of quantum algorithms and
quantum information processing. It has even been argued that performing
experiments in physics can be viewed as simply playing a \textquotedblleft
game" against nature in which the observer tries to maximize the information
obtained from the system under consideration. Eventually, such studies may
even shed light on the great divide between classical and quantum physics
\cite{Lee2002}.

The Prisoner's Dilemma is a widely known example in classical game theory.
It is a two players non-zero sum game where the players may benefit from
unknowing cooperation. Due to the interesting nature of the game and the
fact that communication is forbidden, defection turns out to be the
unilateral best strategy, making it a \emph{Nash}-equilibrium \cite%
{Myerson1991}. The dilemma arises because this strategy does not
provide both players with the collective best payoff (which would be
cooperation). However, extending the game into the quantum domain
resolves the dilemma, as was first pointed out by Eisert \emph{et
al.} \cite{Eisert1999}. In the quantum version of the game,
entanglement introduces some sort of cooperativity between the
players and changes the Nash-equilibrium, so that the collective
best choice for both players and the best individual choices are
equal.

The quantum version of the Prisoner's Dilemma has recently been
experimentally demonstrated using a nuclear magnetic resonance (NMR) quantum
computer \cite{Du2002}. Here we present the first optical implementation
within the one-way model of quantum computation. By employing an all-optical
system where the qubits are encoded in the polarization degree of freedom of
the photons, the quantum states are subject to negligible decoherence and
can easily be distributed among distant players. Moreover, in stark contrast
to NMR quantum computing \cite{Braunstein1999}, in an all-optical
implementation the observed entanglement can always be described as pure -
and since the introduction of entanglement gives rise to the interesting
features of quantum games we consider it important to report on an
experimental realization which is free of any ambiguity in this respect.

Our implementation of the quantum version of the Prisoner's Dilemma follows
a recent proposal \cite{Paternostro2005} which uses optical cluster states
to realize the quantum game's circuit. Since cluster states are the resource
states for one-way quantum computing \cite%
{Briegel01,Nielsen04b,Raussendorf01,Raussendorf03}, our demonstration is
equivalent to playing the game on a quantum computer. The choice of a
photonic system guarantees the externally-controlled implementation of the
player's strategy to a high degree. Additionally, the underlying principles
of one-way quantum computing along with demonstrations of simple quantum
algorithms \cite{Walther05,Prevedel2007,Tame2006} as well as the generation
of cluster states \cite{Kiesel2005a,Zhang2006a,Lu2007} have recently been
successfully demonstrated using linear optics.

The subsequent parts of the paper are structured as follows. A brief
explanation of the Prisoner's Dilemma in the classical as well as in the
quantum domain is given in Section \ref{ch_Pris}. A succinct introduction
into the paradigm of one-way quantum computing and the formulation of the
game in its context follows in Section \ref{ch_OWQC}. The description of our
experimental demonstration as well as the results of our investigation can
be found in Section \ref{ch_exp} while the concluding discussion is in
Section \ref{ch_disc}.

\section{The Prisoner's Dilemma}

\label{ch_Pris}

The Prisoner's Dilemma is a non zero-sum two players game. In the classical
version, each player $j\in \{A,B\}$ independently chooses a strategy $s_{j}$
which is a binary choice $s_{j}\in \{d,c\}$ . The choices are sent to a
supervising referee who computes the payoff of each player $%
\$_{j}(s_{A},s_{B})$ according to a payoff table. Since both players aim to
maximize their individual payoff, the game is known to have a
non-cooperative and selfish character.

The payoff table for player A is shown in Table \ref{payoff_table} and as it
is a symmetric game, player B's payoffs are given by the transposed table.
With the strategy profile $(d,d)$ neither player can increase his/her
individual payoff regardless of the opposition, making it a \emph{Nash
equilibrium} \cite{Myerson1991}. However the cooperative profile $s=(c,c)$
is \emph{Pareto-optimal} \cite{Fudenberg1991} since no player can increase
their payoff by changing strategy, without reducing the payoff of the
opponent. Classically, the Dilemma arises since $(d,d)$ is a dominant
profile (rational reasoning causes both players to choose this strategy) but
the associated payoff is not the overall best available to them.


\begin{table}[h]
\centering%
\begin{tabular}{|c||c|c|}
\hline
A$\backslash $B & $c$ & $d$ \\ \hline\hline
$c$ & $\$_{A}(c,c)=3$ & $\$_{A}(c,d)=0$ \\ \hline
$d$ & $\$_{A}(d,c)=5$ & $\$_{A}(d,d)=1$ \\ \hline
\end{tabular}%
\caption{Payoff table of player A for the classical Prisoner's
Dilemma. Since this is a symmetric game, player B's payoffs are
given by the transposed table.} \label{payoff_table}
\end{table}

\begin{figure}[tbp]
\centering
\includegraphics[width=10 cm]{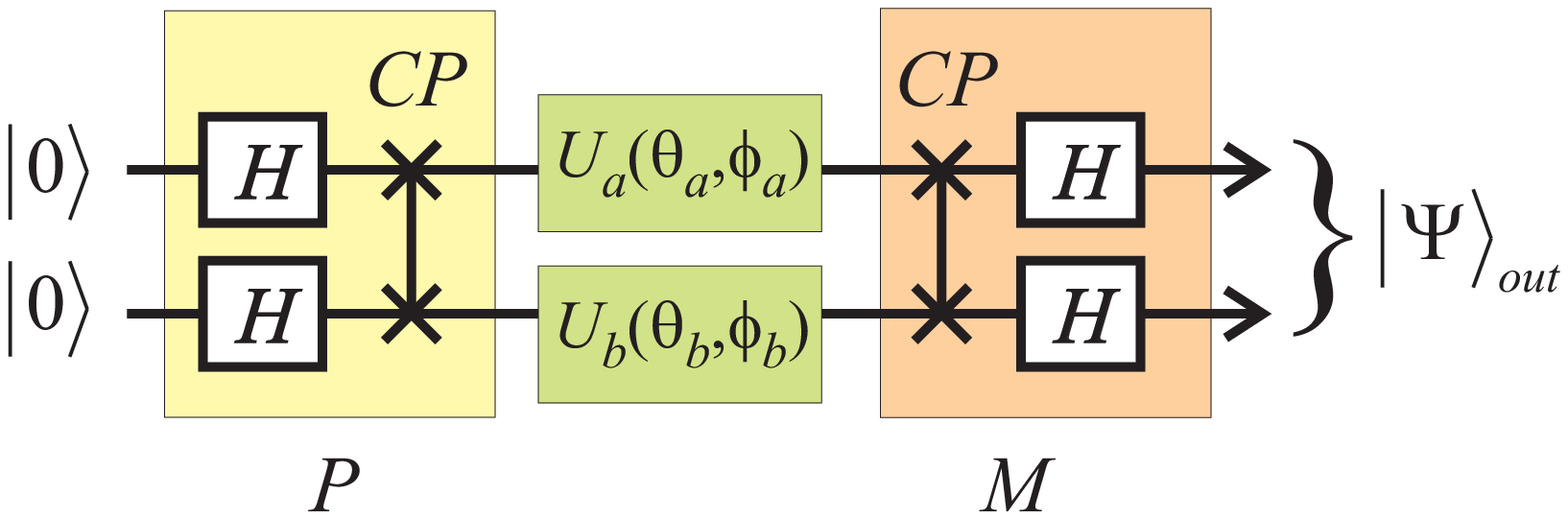}
\caption{Quantum circuit of a two player quantum game, H is a Hadamard gate
and $CP$ a CPhase gate. The output state of the circuit is sent to a referee
who computes the payoffs.}
\label{qcircuit}
\end{figure}

In the quantum version of this game, however, this dilemma can be solved.
Introducing entanglement provides both players with the ability to cooperate
and therefore with an increased strategy space, effectively changing the
Nash-equilibrium \cite{Eisert1999}. Suppose the strategy is realized by
qubits, on which each player can perform their strategy by applying unitary
operations. Following \cite{Eisert1999}, the new strategy space is spanned
by the unitary operator
\begin{equation}
U_{j}\left( \theta _{j},\phi _{j}\right) =\left(
\begin{array}{cc}
e^{-i\phi _{j}}\cos (\theta _{j}/2) & -\sin (\theta _{j}/2) \\
\sin (\theta _{j}/2) & e^{i\phi _{j}}\cos (\theta _{j}/2)%
\end{array}%
\right) ,  \label{unit_op}
\end{equation}%
where $\theta \epsilon \lbrack 0,\pi ]$ and $\phi \epsilon \lbrack 0,\pi /2]$%
. The respective classical strategies $c$ and $d$ are realized by $%
U_{j}(0,0) $ and $U_{j}(\pi ,0)$. Before and after the operation of
the players, the two qubits are subjected to entangling operations
denoted $P$ and $M$ (see Fig. \ref{qcircuit}), which in our specific
game, are a combination of Hadamard and CPhase ($CP$) operations (a
CPhase operation is a two-qubit entangling gate, which in the
logical basis adds a $\Pi$  phase shift to the $|11\rangle$ term).
Without those entangling steps the quantum version would not differ
from a probabilistic, classic game. The corresponding quantum
circuit is shown in Fig. \ref{qcircuit}. To compute the payoffs in
the quantum version the referee projects the two qubit state onto
the computational basis $\{|0\rangle,|1\rangle\}$ and distributes
the payoff according to the payoff table.

Depending on the player's actual choice of strategy (i.e. the unitary $U_{j}$%
), the cooperativity due to the shared entanglement is preserved, giving
rise to a Pareto optimal point that coincides with the Nash-equilibrium.

\section{Playing the game on a one-way quantum computer}

\label{ch_OWQC}

The entangling stages ($P$ and $M$) that are introduced in the quantum
version of the game can be engineered by two-qubit gates. Two-qubit gates
are crucial elementary gates for quantum computation \cite{Barenco95} and
have recently been demonstrated in the all-optical regime \cite%
{OBrien03,Gasparoni04,Pittman03,Langford2005,Kiesel2005,Okamoto2005}. In the
one-way model of computation, such gates can be implemented by a proper
measurement pattern on a sufficiently large entangled resource state
(cluster state) \cite{Raussendorf01,Walther05}. A specific way to implement
the Prisoner's Dilemma on an all-optical one-way quantum computer was
proposed by Paternostro \emph{et al.} \cite{Paternostro2005}. The main
advantage of the one-way model is that the entangling gates are already
intrinsically implemented in the structure of the cluster state, such that
the actual game can easily be carried out by single-qubit rotations only. We
will briefly discuss this in the following.

In the alternative and elegant model of one-way quantum computing the
information processing is achieved by performing single-qubit measurements
on a highly-entangled multi-particle cluster state \cite{Raussendorf01}.
This shifts the difficulty of realizing unitary gates to the generation of
an appropriately designed multi-particle entangled state - often called
cluster state - which serves as a resource for the computation. The
processing of information is accomplished by sequential single-qubit
measurements on the cluster qubits, greatly facilitating the computation
itself. Given a cluster state, measurements in the computational basis $%
\{|0\rangle,|1\rangle\}$ have the effects of disentangling the qubit, while
leaving the remaining qubits entangled. Measurements performed in a
different basis denoted $\{|\alpha_{+}\rangle,|\alpha_{-}\rangle\}$, where $%
|\alpha_{\pm}\rangle=(|0\rangle\pm e^{ i\alpha}|1\rangle)/\sqrt{2}$, also
effectively rotate the logical qubit that undergoes the computation. In our
case, the rotation is around the z-axis $R_{z}(\alpha)=exp(i\alpha%
\sigma_{z}/2)$ and followed by a Hadamard gate $H$. Rotations around the
x-axis, i.e. $R_{x}(\alpha)=exp(i\alpha\sigma_{x}/2)$ can be implemented
through the matrix identity $R_{x}(\alpha)=HR_{z}(\alpha)H$. An elaborated
and detailed introduction to experimental one-way quantum computing is
discussed in~\cite{Walther05,Prevedel2007}. Any complex operation
(consisting of one- and two-qubit gates) can be carried out by a suitable
choice of measurement patterns on a sufficiently large cluster state, so
that, literally, the specific sequence of measurements forms the algorithms
that is computed.

\begin{figure}[tbp]
\centering
\includegraphics[width=10 cm]{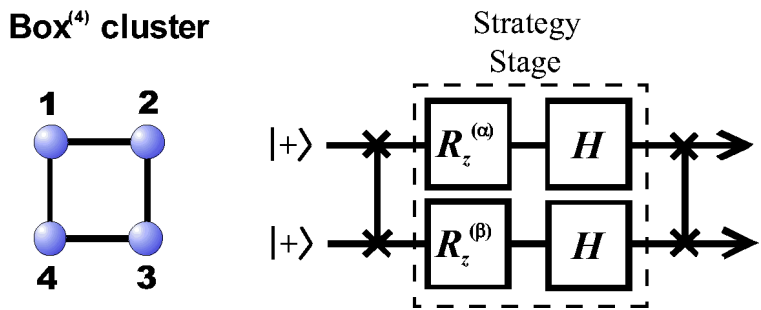}
\caption{Left: Schematic representation of a box-cluster state. Physical
qubits (blue spheres) are entangled to their nearest neighbors (indicated by
a black line) by applying CPhase gates between them. Right: The quantum
circuit realized by the box-cluster state. Note that the input states are
initialized as the logical $\left\vert +\right\rangle$ state, which is
equivalent to a Hadamard gate acting on the $\left\vert 0\right\rangle$
state. Therefore the box-cluster implements, up to single-qubit rotations,
the desired quantum circuit depicted in Fig. \protect\ref{qcircuit}.}
\label{box}
\end{figure}

A special cluster state configuration, the box-cluster, is depicted in Fig. %
\ref{box}. It allows the implementation of a given set of unitaries $U_{j}$
on two logical qubits as defined in the quantum circuit in Fig. \ref%
{qcircuit}, by measurement of qubits 1 and 4 in appropriate basis. This
processes the input states, which are initialized as the logical $\left\vert
+\right\rangle $ states, and transfers them across the cluster to qubits 2
and 3. During this process, which is often referred to as \emph{one-bit
teleportation}, the logical qubit undergoes the unitary $U_{j}$, depending
on the measurement basis and its outcome.

However, closer investigation reveals that measurements performed in the $%
\{|\alpha _{+}\rangle ,|\alpha _{-}\rangle \}$ basis would only allow $%
R_{z}(\alpha )H$ operations, which do not belong to the strategy
space defined by Eq.(\ref{unit_op}) apart for $\alpha =\pi /2$,
consequently limiting the strategy space to $\{c,q\}$ where
$q=U_{j}\left( 0,\pi /2\right) $. Therefore we have to introduce an
additional single-qubit rotation before the measurements, as
described in \cite{Paternostro2005}. Then the strategy space can be
increased to $\{c,d,q\left( \alpha \right) \}$ where $q\left( \alpha
\right) =U_{j}\left( 0,\alpha \right) $, which allows an
experimental realization of the quantum version of the game, as will
be discussed in the following section.

\section{Experimental Realization}

\label{ch_exp}

\begin{figure}[tbp]
\centering
\includegraphics[width=16 cm]{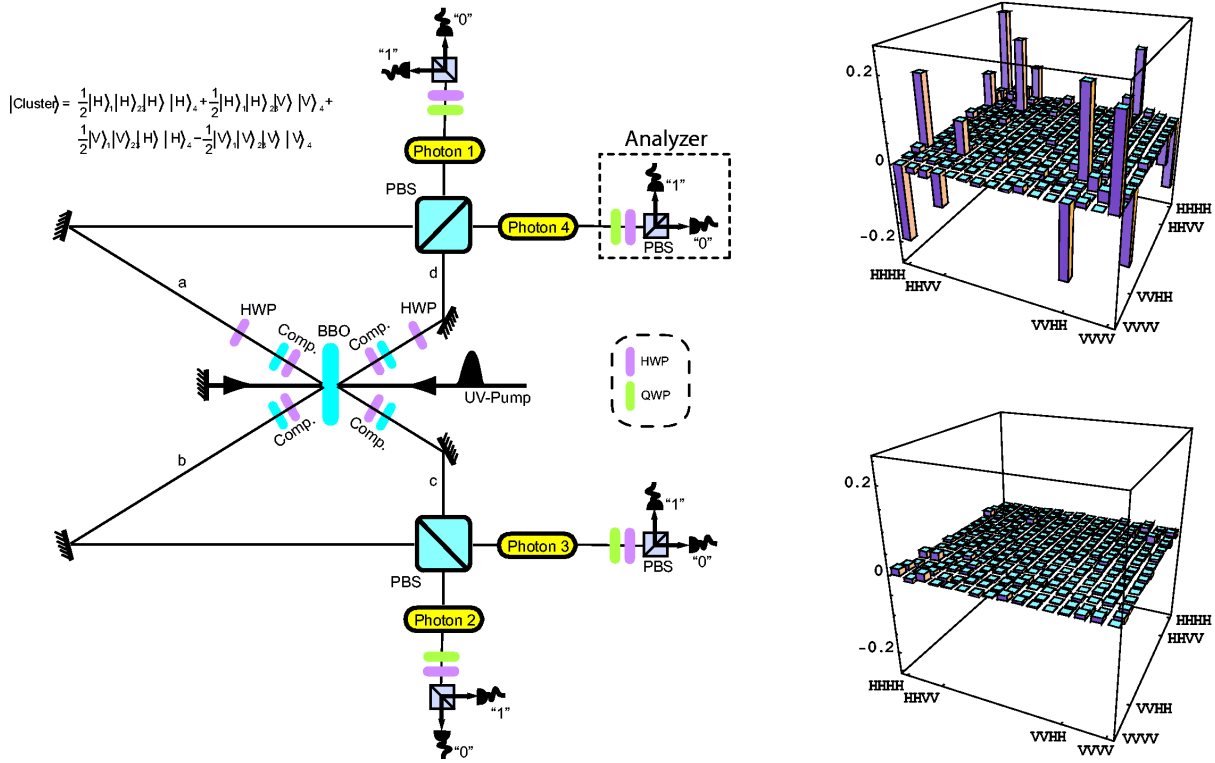}
\caption{Left: Schematic drawing of the experimental setup that is
employed to realize the quantum version of the Prisoner's Dilemma.
Whenever one photon is emitted into each of the four output ports of
the PBSs (mixing modes $a-d$ and $b-c$), a photonic 4-qubit cluster
state is generated. The analyzers, which consist of QWP, HWP and
PBS, allow measurements in an arbitrary polarization basis and
therefore the implementation of the quantum game. Details are
discussed in the text. Right: Tomographic plot of the generated
cluster state with the real part (upper plot) and imaginary part
(lower plot) of the density matrix.} \label{setup}
\end{figure}

The cluster state creation is based on a interferometric method employing
entangled photon pairs produced by spontaneous parametric down-conversion
\cite{Kwiat95} and was first demonstrated in Ref. \cite{Walther05}. An
ultra-violet laser pulse ($1$ W, $150$ fs, $\lambda=394.5$ nm) passes twice
through a non-linear crystal (BBO), thereby generating
polarization-entangled photon pairs in both the forward (modes $a$ and $b$)
and backward (modes $c$ and $d$) direction (see Fig. \ref{setup}). Half-wave
plates (HWP) and BBO crystals are used to counteract walk-off effects in the
down-conversion crystal \cite{Kwiat95}. They are aligned such that $\Phi^{-}$
and $\Phi^{+}$ states are emitted in the forward and backward direction,
respectively. Taking into account the possibility of double-pair emission
into each direction and the action of the polarizing beam splitters (PBS)
mixing modes $a-d$ and $b-c$, the four amplitudes of the cluster state can
be generated by rotating an additional HWP in mode $a$ (see Ref. \cite%
{Walther05} for further details). Subsequently, the photons pass narrowband
interference filters ($\delta\lambda=3nm$), and are then coupled into
single-mode fibres and guided to the detection stage, where the photon's
polarization is detected in an arbitrary basis using a combination of
quarter-wave plates (QWP), HWP and PBS (see Fig. \ref{setup}). A
multichannel coincidence unit allows simultaneous detection of all relevant
16 four-fold coincidence events, therefore significantly speeding up the
tomography process. The relative phase between the forward and backward
emission in the setup sets the phases of the four individual terms of the
cluster state. In the experiment, this is achieved with a piezo actuator
translating the pump mirror. In the experiment, generation of the cluster
state is retrodictive: it is known to have been prepared when one photon in
each output port of the PBS's is detected. This postselection technique is
well established in linear optics and ensures that photon loss and
photodetector inefficiency do not affect the experimental results.

In an ideal case, the following four-photon state is produced by the
experimental set-up:
\begin{equation}  \label{Mcluster}
\left\vert \Phi_{c}\right\rangle=\frac{1}{2}(\left\vert
0000\right\rangle+\left\vert 0011\right\rangle+\left\vert
1100\right\rangle-\left\vert 1111\right\rangle)_{1234}
\end{equation}
with $\left\vert 0\right\rangle_{j}$ ($\left\vert 1\right\rangle_{j}$)
embodied by the horizontal (vertical) polarization state of one photon
populating a spatial mode $j=1,..,4$. The state $\left\vert
\Phi_{c}\right\rangle$ can be converted to the box cluster state (Fig. \ref%
{box}) by the local unitary operation $H_{1}\otimes H_{2}\otimes
H_{3}\otimes H_{4}$ and a swap (or relabeling) of qubits 2 and 3 \cite%
{Walther05}.

The quality of the generated cluster state is quantified by performing full
quantum state tomography \cite{James01}. The reconstructed density matrix of
the experimentally produced state, $\varrho $, is presented in Fig. \ref%
{setup} and has a fidelity with the ideal state in Eq.~(\ref{Mcluster}) of $%
F=\left\langle \Phi _{c}|\varrho |\Phi _{c}\right\rangle =0.62\pm
{0.01}$. The error bar of this result was estimated by performing a
100 run Monte Carlo simulation of the whole state tomography
analysis, with Poissonian noise added to the count statistics in
each run~\cite{Langford2005}. Higher fidelities are difficult to
achieve due to phase instability during the lengthy process of state
tomography and non-ideal optical elements employed in the setup.
However, it is well-above the limit $F=0.5$ for any biseparable
four-qubit state~\cite{Toth2005}. This demonstrates the presence of
genuine four particle entanglement and confirms that such an
experimental state can be used for the quantum protocol under
consideration.

Starting from the state (\ref{Mcluster}) the game is implemented by
projecting the photons 1 and 4 onto the state $|\theta
_{1,4}\rangle_{1,4}=\cos \left( \theta _{1,4}\right) \left\vert
0\right\rangle _{1,4}+\sin \left( \theta _{1,4}\right) \left\vert
1\right\rangle _{1,4}$ resulting in the state $|\psi \left(
\theta_{1};\theta _{4}\right) \rangle _{23}=\,_{1}\langle
\theta|_{4}\langle \theta _{4}|\Phi _{c}\rangle _{1234}$ where
$\theta_{1}$  and  $\theta_{4}$  determine the strategies of players
A and B, respectively, up to a rotation on the remaining photons.
This projection in the laboratory basis is equivalent, up to a
Hadamard rotation, to the box cluster state. The final state
$\left\vert \Psi \right\rangle _{23}^{out}$, after the projection
and any relevant $\sigma _{y}$ operations are applied to them,
resides on qubits 2 and 3 which are sent to the referee who
calculates the payoff. The experimental parameters for the chosen
strategies can be inferred from Table \ref{exp_para}. In the
appendix we give a detailed derivation of this table.

\begin{table}[h]
\centering%
\begin{tabular}{|c|c|c|c|}
\hline
A $\backslash $ B & $c$ & $d$ & $q\left( \alpha _{B}\right) $ \\ \hline
$c$ & $\mathbb{I}\otimes \mathbb{I} \left\vert \psi \left( 0;0\right)
\right\rangle $ & $\sigma _{y}\otimes \mathbb{I}\left\vert \psi \left(
0;-\pi /2\right) \right\rangle $ & $\mathbb{I}\otimes \mathbb{I}\left\vert
\psi \left( 0;\alpha _{B}\right) \right\rangle $ \\ \hline
$d$ & $\mathbb{I}\otimes \sigma _{y}\left\vert \psi \left( -\pi /2;0\right)
\right\rangle $ & $\sigma _{y}\otimes \sigma _{y}\left\vert \psi \left( -\pi
/2;-\pi /2\right) \right\rangle $ & $\mathbb{I}\otimes \sigma _{y}\left\vert
\psi \left( -\pi /2;\alpha _{B}\right) \right\rangle $ \\ \hline
$q\left( \alpha _{A}\right) $ & $\mathbb{I}\otimes \mathbb{I}\left\vert \psi
\left( \alpha _{A};0\right) \right\rangle $ & $\sigma _{y}\otimes \mathbb{I}%
\left\vert \psi \left( \alpha _{A};-\pi /2\right) \right\rangle $ & $\mathbb{%
I}\otimes \mathbb{I}\left\vert \psi \left( \alpha _{A};\alpha _{B}\right)
\right\rangle $ \\ \hline
\end{tabular}
\caption{Table of the states after the players implemented their strategies.
$\left\vert \protect\psi \left( \protect\alpha _{A},\protect\alpha %
_{B}\right) \right\rangle $ is the state after both player applied their
projections with angles $\protect\alpha _{A}$ and $\protect\alpha _{B}$ as
described in the text. The final state $\left\vert \Psi \right\rangle
_{23}^{out}$ is obtained by applying an additional rotation $\protect\sigma %
_{y}$ if necessary.}
\label{exp_para}
\end{table}

\begin{figure}[tbp]
\centering
\includegraphics[width=14 cm]{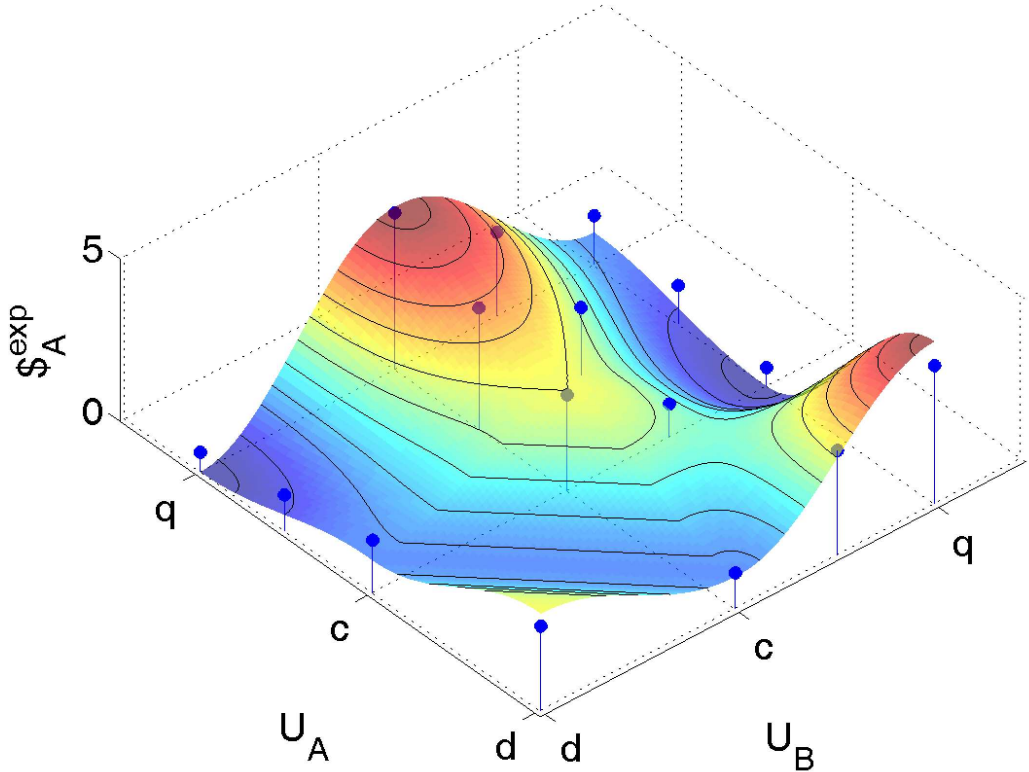}
\caption{Graphical representation of the theoretical (surface) and
measured (dots) payoffs of player A as a function of both players'
strategies. The interval [d,c] is defined by the strategies
$U_{j}\left( \protect\theta,0 \right) $ with $\protect\theta\in
[\protect\pi,0]$ and [c,q] by the
strategies $U_{j}\left( 0,\protect\phi \right) $ with $\protect\phi\in [0,%
\protect\pi/2]$. The strategy profile (d,d) is Pareto-optimal and a
Nash equilibrium thus resolving the dilemma occuring in the
classical version of the game.} \label{results}
\end{figure}

Experimentally the payoffs are determined as follows. We project the
remaining two photons onto the $\{|0\rangle ,|1\rangle \}$ basis and measure
the probabilities $p_{ij}=\left\vert \left\langle ij\right\vert \left. \Psi
\right\rangle _{23}^{out}\right\vert ^{2}$. The payoff of player A is then
computed using

\begin{equation}
\$_{A}^{\exp
}(s_{A},s_{B})=\$_{A}(c,c)p_{00}+\$_{A}(c,d)p_{01}+\$_{A}(d,c)p_{10}+%
\$_{A}(d,d)p_{11}
\end{equation}

For each player we have chosen the following 4 strategies $\left\{
c,d,q\left( \pi /4\right) ,q\left( \pi /2\right) \right\} $. Fig. \ref%
{results} shows the experimental payoffs for all possible combinations of
the implemented strategies. For comparison, the expected, ideal payoff
function is also shown as a surface plot. We find good agreement between the
measured and expected values. The discrepancies are due to the non-ideal
cluster state resource at hand. Unwanted correlations are known to affect
the computation performed according to the one-way model in a
protocol-dependent fashion \cite{Tame2006a}. Moreover, some of the payoffs
corresponding to specific strategic moves played by A and B, suffer from the
imperfect resource more than other, due to the specific nature of the
measurement being performed. We emphasize that although we cannot implement $%
U\left( \alpha ,0\right) $ strategies with arbitrary $\alpha $, our strategy
space is still large enough to resolve the dilemma.

\section{Discussion and outlook}

\label{ch_disc}

We have experimentally demonstrated the application of a
measurement-based protocol to realize a quantum version of the
Prisoner's Dilemma. Our implementation is based on entangled
photonic cluster states and constitutes the first realization of a
quantum game in the context of one-way quantum computing.
Furthermore, our particular realization is especially suited for
playing between distant parties. Because all the entangling
operations preparing the cluster state are done locally by the
referee, it is easy to distribute the entangled photons, even over
large distances. Here we note that, of course, the game can also be
played using an ancillary entangled pair for the realization of the
disentangling CPhase gate. In this scenario, initially both players
share one particle of an entangled photon pair and apply a
polarization rotation on their respective photon $U_{j}\left( \theta
_{j},\phi _{j}\right) $ (corresponding to their chosen strategy).
The photons are then sent to the referee who applies the
disentangling operation with an ancillary, entangled pair
\cite{Gasparoni04}. However such an operation experimentally
requires interferometric stability between the initial and the
ancilla pairs, a very difficult experimental challenge if the
players reside at distant locations.

Another interesting feature is that, in our demonstration, the entanglement
generation is decoupled from the actual processing of the quantum mechanical
information. It remains an open question whether applications of few qubit
cluster states could facilitate some kind of remote quantum information
processing, e.g. multi-party quantum communication protocols \cite{Munro2005}%
. Nonetheless, we also expect that the simple nature of our demonstration
will trigger further interest in the one-way model of quantum computation,
in particular in the realization of simple quantum algorithms.

\ack
We want to thank M. Paternostro and M.S. Tame for fruitful
discussions as well as A. Fedrizzi for careful reading of the
manuscript. We acknowledge financial support by the FWF, the
European Commission under the Integrated Project Qubit Applications
(QAP) and the U.S. Army Research Office funded DTO (QCCM).

\appendix
\setcounter{section}{1}

\section*{Appendix}

In order to find the correspondence between the quantum circuit which
describes the game, as depicted in Fig.~\ref{qcircuit}, and the sequence of
measurements on the cluster state, we compare the output state of the
circuit for each chosen strategy to the corresponding output state of the
one-way computation sequence. The output state of the circuit for the input
state $\left\vert 00\right\rangle $ is

\[
\centering \left\vert \Psi\right\rangle_{out} =\left[ H\otimes H\right]
\cdot CP\cdot \left[ U_{a}\left( \theta _{a},\phi _{a}\right) \otimes
U_{b}\left( \theta _{b},\phi _{b}\right) \right] \cdot CP\cdot \left[
H\otimes H\right] \left\vert 00\right\rangle
\]
where the CPhase gate ($CP$) is defined by
\[
\centering CP=\left(
\begin{array}{cccc}
1 & 0 & 0 & 0 \\
0 & 1 & 0 & 0 \\
0 & 0 & 1 & 0 \\
0 & 0 & 0 & -1%
\end{array}%
\right) .
\]

Table \ref{output_circuit} shows the output states as a function of
the player's local strategies. Projecting these states onto the
computational basis leads to the payoffs shown in Table
\ref{payoff}. When the (dis)entangling operations are removed from
the circuit this payoff table reduces to the original Table
\ref{payoff_table}.

\begin{table}[h]
\centering%
\begin{tabular}{|c||c|c|c|}
\hline
A$\backslash $B & $c$ & $d$ & $q\left( \alpha _{B}\right) $ \\ \hline\hline
$c$ & $\left\vert 00\right\rangle $ & -$\left\vert 11\right\rangle $ & $\cos
\left( \alpha _{B}\right) \left\vert 00\right\rangle -i\sin \left( \alpha
_{B}\right) \left\vert 01\right\rangle $ \\ \hline
$d$ & -$\left\vert 11\right\rangle $ & -$\left\vert 00\right\rangle $ & $%
i\sin \left( \alpha _{B}\right) \left\vert 10\right\rangle -\cos \left(
\alpha _{B}\right) \left\vert 11\right\rangle $ \\ \hline
$q\left( \alpha _{A}\right) $ & $%
\begin{array}{c}
\cos \left( \alpha _{A}\right) \left\vert 00\right\rangle \\
-i\sin \left( \alpha _{A}\right) \left\vert 10\right\rangle%
\end{array}%
$ & $%
\begin{array}{c}
i\sin \left( \alpha _{A}\right) \left\vert 01\right\rangle \\
-\cos \left( \alpha _{A}\right) \left\vert 11\right\rangle%
\end{array}%
$ & $%
\begin{array}{c}
\cos \left( \alpha _{A}\right) \cos \left( \alpha _{B}\right) \left\vert
00\right\rangle -i\cos \left( \alpha _{A}\right) \sin \left( \alpha
_{B}\right) \left\vert 01\right\rangle \\
-i\sin \left( \alpha _{A}\right) \cos \left( \alpha _{B}\right) \left\vert
10\right\rangle -\sin \left( \alpha _{A}\right) \sin \left( \alpha
_{B}\right) \left\vert 11\right\rangle%
\end{array}%
$ \\ \hline
\end{tabular}%
\caption{Output states from the game circuit as a function of players' A and
B strategies. Although these states are separable, they cannot be obtained
by local unitary operations and without the action of (dis-)entangling
operations between the players.}
\label{output_circuit}
\end{table}

\begin{table}[h]
\centering%
\begin{tabular}{|c||c|c|c|}
\hline
A$\backslash $B & $c$ & $d$ & $q\left( \alpha _{B}\right) $ \\ \hline\hline
$c$ & $\$_{A}(c,c)$ & $\$_{A}(d,d)$ & $\left\vert \cos \left( \alpha
_{B}\right) \right\vert ^{2}\$_{A}(c,c)+\left\vert \sin \left( \alpha
_{B}\right) \right\vert ^{2}\$_{A}(c,d)$ \\ \hline
$d$ & $\$_{A}(d,d)$ & $\$_{A}(c,c)$ & $\left\vert \sin \left( \alpha
_{B}\right) \right\vert ^{2}\$_{A}(d,c)+\left\vert \cos \left( \alpha
_{B}\right) \right\vert ^{2}\$_{A}(d,d)$ \\ \hline
$q\left( \alpha _{A}\right) $ & $%
\begin{array}{c}
\left\vert \cos \left( \alpha _{A}\right) \right\vert ^{2}\$_{A}(c,c) \\
+\left\vert \sin \left( \alpha _{A}\right) \right\vert ^{2}\$_{A}(d,c)%
\end{array}%
$ & $%
\begin{array}{c}
\left\vert \sin \left( \alpha _{A}\right) \right\vert ^{2}\$_{A}(c,d) \\
+\left\vert \cos \left( \alpha _{A}\right) \right\vert ^{2}\$_{A}(d,d)%
\end{array}%
$ & $%
\begin{array}{c}
\left\vert \cos \left( \alpha _{A}\right) \cos \left( \alpha _{B}\right)
\right\vert ^{2}\$_{A}(c,c) \\
+\left\vert \cos \left( \alpha _{A}\right) \sin \left( \alpha _{B}\right)
\right\vert ^{2}\$_{A}(c,d) \\
+\left\vert \sin \left( \alpha _{A}\right) \cos \left( \alpha _{B}\right)
\right\vert ^{2}\$_{A}(d,c) \\
+\left\vert \sin \left( \alpha _{A}\right) \sin \left( \alpha _{B}\right)
\right\vert ^{2}\$_{A}(d,d)%
\end{array}%
$ \\ \hline
\end{tabular}%
\caption{Payoffs for player A computed using the states from Table \protect
\ref{output_circuit}.}
\label{payoff}
\end{table}

Next we show how a cluster state can be used to simulate the quantum
circuit corresponding to the quantum game. The cluster state
$\left\vert \Phi_{c} \right\rangle=\frac{1}{2}(\left\vert
0000\right\rangle +\left\vert 0011\right\rangle +\left\vert
1100\right\rangle -\left\vert 1111\right\rangle )_{1234}$ is
projected onto a two photon state by projecting the qubit 1 and 4
onto the states $\cos \left( \theta _{1,4}\right) \left\vert
0\right\rangle _{1,4}+e^{i\varphi _{1,4}}\sin \left( \theta
_{1,4}\right) \left\vert 1\right\rangle _{1,4}$. We verify that the
remaining two-photon state is equivalent to the circuit outcome up
to a local rotation on each remaining qubit. Before the rotation the
state is

\begin{eqnarray*}
\left\vert \psi \left( \theta _{1},\varphi _{1};\theta _{4},\varphi
_{4}\right) \right\rangle _{23} &=&\cos \left( \theta _{1}\right) \cos
\left( \theta _{4}\right) \left\vert 00\right\rangle _{23}+e^{i\varphi
_{4}}\cos \left( \theta _{1}\right) \sin \left( \theta _{4}\right)
\left\vert 01\right\rangle 
_{23} \\
&+&e^{i\varphi _{1}}\sin \left( \theta _{1}\right) \cos \left(
\theta _{4}\right) \left\vert 10\right\rangle _{23}-e^{i(\varphi
_{1}+\varphi _{4})}\sin \left( \theta _{1}\right) \sin \left( \theta
_{4}\right) \left\vert 11\right\rangle _{23}
\end{eqnarray*}

When Player A and B apply a rotation $R_{j}\left( \alpha _{j},\beta
_{j},\gamma _{j}\right) $ on qubit 3 and qubit 2 respectively, the final
output state is

\[
\left\vert \Psi \right\rangle _{23}^{out}=R_{B}\left( \alpha _{B},\beta
_{B},\gamma _{B}\right) \otimes R_{A}\left( \alpha _{A},\beta _{A},\gamma
_{A}\right) \left\vert \psi \left( \theta _{1},\varphi _{1};\theta
_{4},\varphi _{4}\right) \right\rangle _{23}
\]%
where
\[
R\left( \alpha ,\beta ,\gamma \right) =R_{z}\left( \alpha \right)
R_{x}\left( \beta \right) R_{z}\left( \gamma \right) =\left(
\begin{array}{cc}
e^{i\left( \alpha -\gamma \right) }\cos \left( \beta /2\right) & -e^{i\left(
\alpha +\gamma \right) }\sin \left( \beta /2\right) \\
e^{i\left( \alpha -\gamma \right) }\sin \left( \beta /2\right) & e^{i\left(
\alpha +\gamma \right) }\cos \left( \beta /2\right)%
\end{array}%
\right)
\]

Table \ref{rot} shows the final states as a function of the strategies.
Although they are not strictly equal to the output of the quantum circuit,
those states lead to the same payoffs when measured in the computational
basis. This proves the equivalence of both approaches and shows that it is
necessary, in order to span the entire strategy space, to extend the cluster
state scheme by allowing arbitrary one-qubit rotations. However, we note
that the strategies $m=U\left( \alpha ,0\right) $ are not accessible because
the output of the circuit for the strategy $(s_{A},s_{B})=\left( q\left(
\alpha \right) ,c\right) $ is $\cos \left( \alpha /2\right) \left\vert
00\right\rangle -\sin \left( \alpha /2\right) \left\vert 11\right\rangle $.
Such an output cannot be achieved using a cluster state of the form of Eq. %
\ref{Mcluster} for any $\alpha $ different from $0$ or $\pi $. A six photon
cluster state \cite{Lu2007} would be required to implement the whole space
of strategies $U_{j}\left( \theta _{j},\phi _{j}\right) $.

\begin{table}[h]
\centering%
\begin{tabular}{|c|c|c|c|}
\hline
A $\backslash $ B & $c$ & $d$ & $q\left( \alpha _{B}\right) $ \\ \hline
$c$ & $\mathbb{I}\otimes \mathbb{I}\left\vert \psi \left( 0;0\right)
\right\rangle $ & $-i\sigma _{y}\otimes \mathbb{I}\left\vert \psi \left(
0;-\pi /2\right) \right\rangle $ & $\mathbb{I}\otimes \mathbb{I}\left\vert
\psi \left( 0;\alpha _{b}\right) \right\rangle $ \\ \hline
$d$ & $-i\cdot \mathbb{I}\otimes \sigma _{y}\left\vert \psi \left( -\pi
/2;0\right) \right\rangle $ & -$\sigma _{y}\otimes \sigma _{y}\left\vert
\psi \left( -\pi /2;-\pi /2\right) \right\rangle $ & -$\mathbb{I}\otimes
i\sigma _{y}\left\vert \psi \left( -\pi /2;\alpha _{B}\right) \right\rangle $
\\ \hline
$q\left( \alpha _{A}\right) $ & $\mathbb{I}\otimes \mathbb{I}\left\vert \psi
\left( \alpha _{A};0\right) \right\rangle $ & $-i\sigma _{y}\otimes \mathbb{I%
}\left\vert \psi \left( \alpha _{A};-\pi /2\right) \right\rangle $ & $%
\mathbb{I}\otimes \mathbb{I}\left\vert \psi \left( \alpha _{A};\alpha
_{B}\right) \right\rangle $ \\ \hline
\end{tabular}
\caption{Table of the projected states and rotation angles corresponding to
different strategies, with $\mathbb{I}=R_{j}\left( 0,0,0\right) $, $-i%
\protect\sigma _{y}=R_{j}\left( 0,\protect\pi ,0\right) $ and $\left\vert
\protect\psi \left( \protect\alpha _{A};\protect\alpha _{B}\right)
\right\rangle =\left\vert \protect\psi \left( \protect\alpha _{A},0;\protect%
\alpha _{B},0\right) \right\rangle _{23}$.}
\label{rot}
\end{table}

\section*{References}

%

\end{document}